 \documentclass[preprint2]{emulateapj}

\usepackage{hyperref}
\usepackage{url}
\usepackage{times}
\usepackage{natbib}
\usepackage{graphicx}
\usepackage{amsmath}
\usepackage{amsfonts}
\usepackage{amssymb}
\usepackage{pdfpages}
\usepackage{import}
\usepackage{bibentry}

\usepackage{color}

\shorttitle{Solar supergranulation}
\shortauthors{Cossette and Rast}

\begin{document}


\title{Supergranulation as the largest buoyantly driven convective scale of the Sun}


\author{Jean-Francois Cossette} 
\affil{Laboratory for Atmospheric and Space Physics, University of Colorado, Boulder, USA}
\author{Mark P. Rast}
\affil{Department of Astrophysical and Planetary Sciences, Laboratory for
Atmospheric and Space Physics, University of Colorado, Boulder, USA}


\begin{abstract}
Supergranulation is characterized by horizontally divergent flows
with typical length scales of 32 Mm in the solar photosphere. Unlike
granulation, the size of which is comparable to both the thickness of
the radiative boundary layer and local scale height of the plasma in
the photosphere, supergranulation does not reflect any obvious length
scale of the solar convection zone. Early suggestions that the depth
of second helium ionization is important are not supported by numerical
simulations. Thus the origin of the solar supergranulation remains
largely a mystery. Moreover, observations of flows in the photosphere 
using either Doppler imaging or correlation or feature tracking show a
monotonic decrease in power at scales larger than supergranulation. Both
local area and global spherical shell simulations of solar convection
by contrast show the opposite, a power law increase in horizontal flow
amplitudes to low wavenumber. Here we examine this disparity, and
investigate how the solar supergranulation may arise as a consequence
of strong photospheric driving and non-local heat transport by cool
diving plumes. Using three dimensional anelastic
simulations with surface driving, we show that the kinetic energy of
largest convective scales in the upper layers of a stratified domain
reflects the depth of transition from strong buoyant driving to adiabatic
stratification below. We interpret the observed monotonic decrease in
solar convective power at scales larger than supergranulation to be a
consequence of this rapid transition, and show how the supergranular
scale can be understood as the largest buoyantly driven mode of convection
in the Sun.
\end{abstract}


\keywords{Sun: interior}

\section{Introduction}
Scales of solar convection fall into three main categories, 
granules, mesogranules and supergranules, with recent observations
hinting at the possibility of giant cells \citep{Hatha13}. Granules
(1Mm diameter, 0.2 hr lifetime) are the signature of convective
cells driven in the highly superadiabatic layers of the photosphere.
Direct observation in continuum intensity images
has confirmed their convective nature via the correlation of vertical
velocity with intensity (e.g. \cite{Nord09}). Supergranules ($32$Mm
diameter; 1.8 day lifetime) are observed largely as a horizontal flow
using either Doppler imaging, magnetic feature or granule tracking, or
local helioseismology \citep{Hana16}. The horizontally divergent motion
and cellular nature of supergranulation suggest a convective origin. The
presence of magnetic flux elements in network boundaries makes direct
observation of the vertical velocity and intensity correlation difficult,
though the temperature contrast across the cells has been measured
~\citep{Gold09}. The physical mechanism responsible for supergranulation
remains unclear. The early suggestion that the second ionization of helium
plays an important role \citep{Leigh62,SimLeigh64,Novem81} is not supported
by numerical simulation~\citep{RastToomre93,Lord14}, while the later
suggestion that supergranulation results from self-organization of granular
flows \citep{Rieutord00,Rast03,Crouch07} may be more relevant on
mesogranular scales~\citep{Cat01,Berrilli05,Leit05,Duval10}. Mesogranules
($5$Mm diameter; 3 hour lifetime) are intermediate scale structures seen
primarily as vertical velocity in time-averaged Doppler maps
\citep{Novem81}. Their existence as a real convective
feature disctinct from both granules and supergranules is still
debated~\citep{Nov89,Berrilli13}. 

Convective structures much larger than supergranules, including
so-called giant cells, are predicted by both mixing length theories
and global models of solar convection \citep{CD1996,Miesch08}.
However, observations suggest that the velocities
associated with these large-scale flow motions are significantly
weaker than predicted. Time distance helioseismology provides the
most severe constraint, with large-scale velocity amplitudes 
at 28-56Mm depth measured to be orders of magnitude smaller than
in models \citep{Hana14,Hana12,Hana10}. However, ring diagram helioseismic
analysis does not confirm this, instead showing at 30Mm depth a
continuous increase of power  to scales larger than supergranulation,
in good agreement with numerical experiments \citep{Greer15}.  Where
models and observations most fundamentally disagree is in the surface
layers. Horizontal velocity power spectra obtained from Doppler
imaging and correlation tracking of flow features at the solar surface
reveal peaks corresponding to granular (angular harmonic degree
$l\thicksim 3500$) and supergranular scales ($l\thicksim 120$),
followed by a monotonic decrease in power to larger scales
~\citep{Hatha00,Roudier12,Hana14,Hatha15}. Only recently has a
possible signature of giant cell convection been detected in the
photosphere by carefully tracking the motions of supergranules
~\citep{Hatha13}. Radiative hydrodynamic and magnetohydrodynamic
local area and global models of solar convection, on the other hand,
all show horizontal power increasing monotonically to large
scales~\citep{Lord14,Hana16}.

This discrepancy between modeled and observed  power may be related
to the difficulties global models have reproducing a solar-like
differential rotation in the parameter regime characteristic of the
solar interior. Models indicate that rotationally constrained giant
cells, which transport angular momentum toward the equator, are
essential in maintaining the prograde equatorial differential rotation 
(e.g. \cite{Miesch08}) observed at the photosphere and in accordance
with the angular velocity profiles inferred from helioseismology
\citep{Thompson03}. These solar-like states are achieved when the
flow is rotationally constrained, when the influence of the Coriolis
force dominates over the flow's inertia, which places an upper limit
on the convective flow speeds. This upper limit is weaker than the flow
amplitudes required to transport the solar luminosity in global
simulations (e.g.  \cite{Hotta14}). Moreover, as numerical diffusivities
are lowered, the flows becomes more turbulent and velocity fields
tend to decorrelate, which can lead to faster convective motions
and retrograde differential rotation (poles rotating faster than
the equator) instead of prograde  \citep{Gastine13,MieschFea15}. 
Models of the Sun's convection can reproduce global scale motions
only if the flux through the domain is reduced or the rotation
rate of the star is increased.

These difficulties suggest that global motions in the Sun are weak
enough to be rotationally constrained, with smaller scales carrying 
the convective flux. This is possible if the Sun maintains
a mean gradient in the deep convection zone that is closer to
adiabatic than that achievable in most simulations and dissipative 
effects are negligible, limiting convective driving below the surface 
and leading to a horizontal velocity spectrum in the photosphere 
consistent with that observed~\citep{Lord14}. 
It implies that the solar supergranulation reflects 
the largest buoyantly driven convective scale of the Sun.

In this paper we use 3D numerical simulations of solar convection to
assess this possibility. We do this by vigorously driving surface
convection in the upper layers while simultaneously achieving a nearly
adiabatic stratification in the interior. We examine the spectra that
result and show that they are dependent on the rate of the transition
to adiabatic stratification. We show that, when the transit time of the
fluid parcels across the convection zone is much shorter than the
diffusion time, the depth over which this
transition takes place depends only on the change in the 
filling factor of the downflows with depth due to stratification.

\section{Model}\label{sec:mod}
We simulate solar hydrodynamic convection by solving the Lipps \& Hemler
\citep{LH82} version of the anelastic Euler equations governing
the evolution of momentum and entropy perturbations in a 
gravitationally stratified fluid: 
\begin{eqnarray}
\frac{D{\bf u}}{Dt}&=&-\nabla \pi' +g\frac{\Theta'}{\Theta_o} {\bf k}
~,\label{mmt} \\
\frac{D\Theta'}{Dt}&=&-{\bf u} \cdot \nabla \Theta_a -\frac{ \Theta'}{\tau}
~,\label{entro} \\
\nabla&\cdot &(\rho_o{\bf u}) = 0 ~.\label{cont}
\end{eqnarray}
Here, ${\bf u}$ represents the fluid velocity, $\Theta\equiv\Theta'+\Theta_a$
is the potential temperature (equivalent to the specific
entropy since $ds=c_p d\ln \Theta$, with $c_p$ the specific heat
at constant pressure), and  $\pi'\equiv p'/\rho_\text{o}$ is the
density-normalized pressure perturbation.  

The {\it reference} state about which the anelastic asymptotic expansion
is constructed is denoted by the subscript `o'. It is taken as isentropic
(i.e. $\Theta_o=\ $constant) and in hydrostatic balance, with 
$g(r)= g_b (r_b/r)^{-2}$ the radially-diminishing magnitude of gravity
acceleration pointing in the negative unit vector direction $\bf k$. The
reference state is constructed using values for temperature $T_b$,
density $\rho_b$ and gravitational acceleration $g_b$ at the base of the
domain obtained from a solar structure model \citep{CD1996}.

Primes in equations (\ref{mmt}) and (\ref{entro}) denote perturbations
with respect to an arbitrarily selected {\it ambient} state (denoted by the
subscript `a'). The ambient state chosen here represents the large scale 
thermodynamic equilibrium structure of the Sun on time scales much longer
than the convective turnover time $\tau_c$. The Newtonian cooling term,
$-{ \Theta'}/{\tau}$, in (\ref{entro}) relaxes the potential temperature
to that of the ambient state over a time scale $\tau>>\tau_c$. Thus the
ambient entropy stratification of the domain is maintained over long time
scales, limiting restratification by the convection and driving motions
in regions of superadiabaticity. This is a common approach in atmospheric
models when addressing evolutionary fluctuations about large scale
equilibria \citep{Smo01,Grab02,Warn07} and has also been employed in
some previous simulations of global solar magneto-convection
\citep{Ghiz10,Rac11,Coss13}.

In detail, we construct the ambient state to be strongly subadiabatic  
in the lower portion of the computational domain
$r_b \le r < r_i$, with $r_i = 0.718R_\odot$, adiabatic in the bulk
$r_i \le r \le r_s$, with the value of $r_s$ varying between runs, 
and superadiabatic above $r_s < r \le r_t$. 
It satisfies the polytropic equations for an ideal gas,
$p_a=K\rho_a^{1+1/m}$, $p_a=\rho_a R T_a$, and $dp_a/dr=-\rho_a g$
with a prescribed polytropic index $m=m(r)$.  In the subadiabatic lower
portion of the domain the polytropic index $m$ decreases linearly from
$m_b=3.0$ at the base to $m_i=m_\text{ad}=3/2$ at $r=r_i$.  It then
remains constant at its adiabatic value through the bulk of the domain,
before being set to a superadiabatic value $m_\text{s}<m_\text{ad}$ above
$r=r_s$. The index $m_s$ thus specifies the level of superadiabaticity
in the region $r > r_s$, with the thickness and superadiabaticity of
the upper region differing between simulation runs. The ambient potential
temperature profile for each run is then 
$\Theta_a \equiv T_a (\rho_b T_b/\rho_a T_a)^{1-1 / \gamma}$. 

The relaxation time to the ambient state is set to $\tau=20$ solar days in
the stably stratified and superadiabatic regions and $\tau=1000$ solar
days in the adiabatic bulk of the domain. It is important to note that,
while we call the bulk of the domain adiabatic, this is a statement about
the ambient state only. The long relaxation time in that portion allows 
the convection to reconfigure the thermodynamic gradients therein. This
allows us to study how the interior of the convection zone (CZ) evolves
when subject to strong surface driving and how this in turn effects the
spectrum of the motions that ensue. As a control experiment, we also
consider a simulation for which the ambient state is subadiabatic in the
lower portion $r_b \le r < r_i$ (as described above) but weakly
superadiabatic across the remainder of the domain $r_i \le r < r_t$ . The
relaxation time in that case is taken to be 20 solar days throughout. 

\begin{figure*} [!tbp]
\begin{center}
 \includegraphics[width=0.99\textwidth]{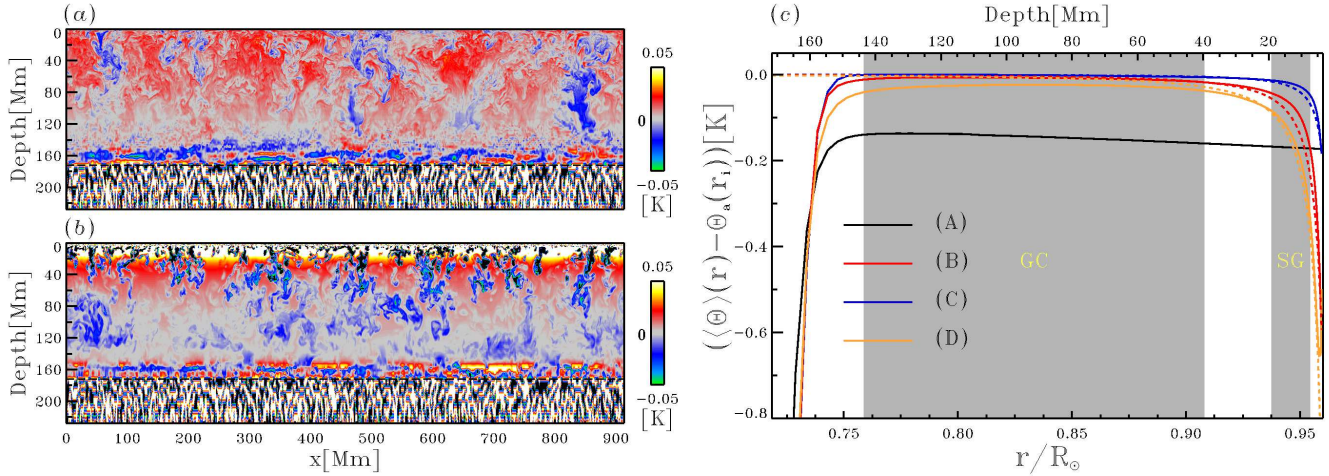}
\caption{Vertical cross-sections of the instantaneous 
deviations $\widetilde{\Theta}\equiv \Theta-\langle\Theta\rangle$
from the horizontal mean of the PT $\langle\Theta\rangle$ taken
from case A---panel (a) and B---panel (b). Case A is
characterized by a weakly superadiabatic ambient profile
($m_s=1.4999998$) across the full depth of the CZ ($r_t \ge r \ge r_i$),
whereas case B  uses a strictly adiabatic profile in the region
($r_s \ge r \ge r_i, r_s\equiv0.96R_\odot$) and a superadiabatic
profile ($m_s=1.4994$) inside a 3.5Mm deep region below the surface
($r_t \ge r \ge r_s$). The horizontal dashed line denotes the location
of the core-enveloppe interface. Low entropy fluid parcels
produced in the driven region pass through the convecting layer
and impact the stable layer below, exciting gravity waves there.
Panel (c) shows profiles of $\langle\Theta(r)\rangle-\Theta_a(r_i)$
for each case in the region $r_s \ge r \ge r_i$ (solid lines). As 
Case B, Cases C and D use, respectively,
strictly adiabatic ambient profiles below $r_s=0.96R_\odot$ and
superadiabatic profiles characterized by $m_s=1.49985$ and 
$m_s=1.4985$ in the region above. Shaded areas labeled `GC' and
`SG' correspond to depth ranges over which
$100\text{Mm}<4H_\rho<300\text{Mm}$ and
$20\text{Mm}< 4H_\rho < 50\text{Mm}$, respectively. The change
in the mean stratification near the
surface in Cases B-D is well reproduced by cold fluid parcels
moving down adiabatically from the height $r_s$ at which they
originate: $\langle \Theta(r)\rangle-\Theta_a(r_i)\approx f(r)\Theta_d$,
where $\Theta_d$ is the parcels' average potential temperature
at $r=r_s$ and $f(r)\equiv f_d \rho_o(r_s)/\rho_o(r)$ is their
filling factor, with $f_d$ the filling factor of downflows at
$r=r_s$ (see dashed lines). The accumulation of low entropy fluid
near the base of the CZ causes $\langle \Theta(r)\rangle$ to decrease
very rapidly in that region.
\label{vmaps}} 
\end{center}
 \end{figure*}

Integration of (\ref{mmt})-(\ref{cont}) is carried out with 
the hydrodynamic solver of the magnetohydrodynamic EULAG
model~\citep{Prusa08, Smolar13}. EULAG employs a two-time-level
flux-form Eulerian non-oscillatory forward-in-time advection
operator \citep{Smolar06}, allowing stable integration of the
equations with all dissipation delegated to the advection scheme's
truncation terms \citep{SmolarPrusa02}. We examine a Cartesian
domain extending from $r_{b}=0.63R_\odot$  to $r_{t}=0.965R_\odot$
in solar radius, which has physical dimensions $910.53\text{Mm}
\times 910.53\text{Mm} \times 227.63 \text{Mm}$ on grids of
$1024^2\times256$ points. The reference states are characterized
by density scale heights $H_\rho=360$km at the surface and $85$Mm
at the base, spanning a total of eleven scale heights across the
domain. Nonuniform gridding in the vertical direction 
accommodates the rapidly decreasing density scale-height near the
top of the domain \citep{Prusa03}. The domain is horizontally
periodic, with vanishing vertical velocity, stress-free horizontal
velocity, and zero flux of the potential temperature imposed at
both upper and lower boundaries.

\section{Results} \label{sec:res}
First we compare two simulations which share approximately the
same convective flux through the bulk of the domain. In Case~A
we specify a weakly superadiabatic ambient state across the full
depth of the layer while in Case B the strongly superadiabatic
ambient state is confined to a 3.5Mm deep region below the surface
(hereafter, the cooling layer). The typical spatial scale of the low
entropy parcels generated in the cooling layer reflects the turbulent
energy injection scale $L\thicksim 4 H_\rho$ in this region
($H_\rho\thicksim 0.36-2.6$Mm) \citep{Rincon07,Lord14}. As can be
seen from Figure~\ref{vmaps}, the flow in Case~A is dominated by
larger scale motions than Case~B. In particular, positive entropy
perturbations in Case A, although weaker, tend to be coherent over
the full depth of the convection zone (Fig.~\ref{vmaps}$a$).

To understand the physical processes shaping the flow structure we
consider additional experiments with different values of the polytropic
index $m_s$ in the cooling layer (Cases C \& D). The mean thermodynamic
stratification (Fig.~\ref{vmaps}$c$) in Case A is characterized by a weakly
superadiabatic mean state ($d\langle \Theta \rangle/dr<0$) throughout.
Cases B-D, on the other hand, show mean states very close to adiabatic
throughout the bulk but strongly superadiabatic near the surface. The 
turbulent energy injection scale in this region is comparable to the
size of supergranules (region SG in Fig.~\ref{vmaps}$c$). The strong
buoyancy force therein thus drives upflows on the scale of supergranulation
(red and yellow areas in Fig.~\ref{vmaps}$b$). The convectively unstable
mean stratification through the bulk of the CZ in Case A (`GC' region
in Fig.~\ref{vmaps}$c$), on the other hand, additionally drives giant
cell scale motions.

It is important to note that, while in Case A the superadiabatic mean entropy
profile is maintained by relaxation to the superadiabatic ambient
state, the relaxation time in Cases B-D is too long to be important in determining
the mean stratification. The strongly superadiabatic region below
the cooling layer is caused by the presence of the cool downflowing
plumes which change the mean state, driving the upflows. This effect
decreases with depth because the filling factor of the downflows
decreases with increasing density until their effect on the mean state
becomes negligible. In all Cases B-D, parcels 
originate from the same depth but have different initial entropy
fluctuations. The downflowing fluid in cases with larger entropy
fluctuations must achieve smaller filling factors before their
influence on the mean state becomes negligible, hence the increase
of the extent of the superadiabatic region when comparing Case C to
B, and B to D. Note that the transit time $\tau_t$ of the cool
plumes across the simulated CZ ($\tau_t\approx 1$ solar day) is shorter than
the time it takes for a parcel to diffuse numerically. As a result,
the change in the mean stratification below the surface is well
approximated by parcels moving adiabatically across the layer (see
dashed lines in Fig.~\ref{vmaps}$c$).  

In the Sun, radiative diffusion dominates over conduction, with
estimates for the radiative diffusivity ranging between
$\kappa\thicksim 10^5-10^7\text{cm}^2\cdot\text{s}^{-1}$
\citep{Miesch05}. The characteristic diffusion timescale
$\tau_d\thicksim \l^2/\kappa$ of a plume with spatial scale
$l\thicksim 300$km (i.e. the thickness of the radiative boundary layer
at the photosphere) is thus between 3-285yr. Assuming that the
transit time $\tau_t$ of cold plumes generated at the photosphere is
of order the turnover time of the largest convective cells
($\thicksim$ 1 month), $\tau_t<<\tau_d$. Solar plumes may then be
expected to behave as in Cases B-D, travelling across the convection
zone without exchanging a significant amount of heat with the
surrounding medium. The consequent superadiabatic mean solar
stratification is due only to the plumes' presence and their
geometry as they move across layers of increasing density.
This implies that the interior stratification of the Sun could be
extremely close to adiabatic with a relatively thin superadiabatic
layer determined by the thermodynamic properties of the granular
downflows in the upper layers.

\begin{figure}[!t]
\begin{center}
 \includegraphics[width=0.49\textwidth]{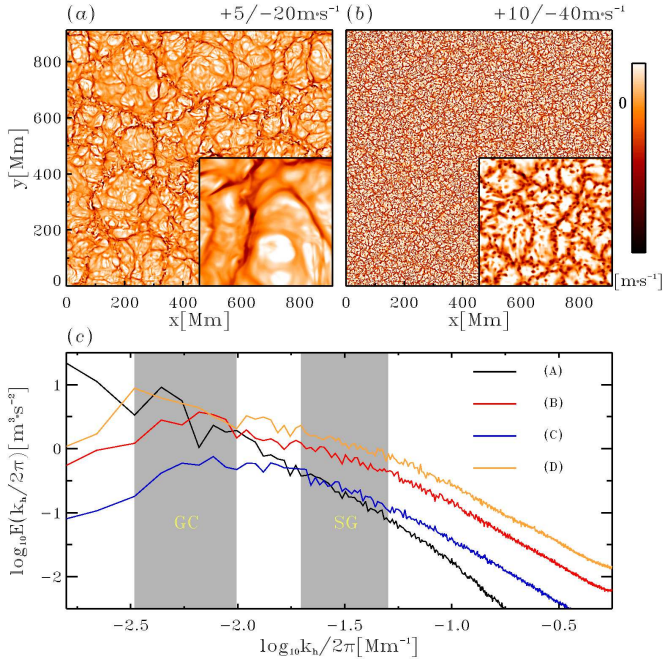}
\caption{Horizontal cross-sections of the instantaneous vertical
velocity $u_r$ taken at 5Mm depth corresponding to Cases
A---panel (a) and B---panel (b). The inserted plot in each panel
shows the magnified view of a $100\text{Mm}^2$ area. As in other
experiments of compressible convection, $u_r$ is characterized by
broad upflows surrounded by a network of narrow downflow lanes
(see \cite{Nord09}, \cite{MieschToomre09} and references therein).
Panel (c): Horizontal velocity power spectra taken at 5Mm depth
for each case. Here, $k_h\equiv2\pi/\lambda$, with $\lambda$ the
horizontal wavelength. Shaded areas labeled `GC' and `SG'
correspond, respectively, to regions where
$100\text{Mm}<\lambda < 300\text{Mm}$ (hereafter,
giant cells) and $20\text{Mm}<\lambda < 50\text{Mm}$ (hereafter,
supergranular scales).\label{hmaps} \vspace{-0.25cm}
} 
\end{center} 
\end{figure}

The characteristic scales of the convective flows reflect 
the depth of the superadiabatic region. Cell diameters in Case A
are much larger than in Case B, with 200Mm scales typical
in Case~A and smaller 40Mm scales in Case~B (Fig.~\ref{hmaps}$a$
and $b$). This difference is reflected in the horizontal
velocity power spectra of the flows (Fig.~2$c$). At 5Mm depth,
the power contained in supergranular scales in Case B exceeds
that of Case A, whereas the opposite is true of the power at
the largest giant-cell scales. 
Increasing the polytropic index $m_s$ in the cooling layer of Case C decreases the power
at all scales relative to Case B, while decreasing it in Case D increases
the amplitude of the convective motions (Fig.~2$c$).  

Notably, the increase of power due to the intensification of convective driving,
when going from Case C to Case B and then from Case B to Case D, is accompanied  
by a corresponding increase of the spatial extent of the superadiabatic region below the
surface (Fig.~\ref{vmaps}$c$). 
The power
contained in giant cells in Cases B-D relative to Case A ($P_{G}/P_{G_A}$)
increases with the superadiabatiticy of the GC region ($\epsilon_\text{G}$), as summarized in 
Table \ref{table}. The table also shows the correlation between the power
contained in the supergranular scales ($P_S/P_{S_A}$) and the
superadiabaticity of the SG region ($\epsilon_\text{S}$).  The
ratio of supergranular to giant cell power ($P_S/P_G$) is the largest
in Case C, where the rate of transition to adiabatic stratification is the
greatest and the characteristic depth (d) of the superadiabatic region
is the smallest. 
\begin{table}[t!]
\begin{center}
 \vspace{+0.5cm}
\begin{tabular}{ccccccc}
Case &  $P_{G}/P_{G_A}$ & $P_{S}/P_{S_A}$ & $P_{S}/P_{G}$ &
$\epsilon_\text{G}$ &
$\epsilon_\text{S}$ & d \\
     & & & & $[\times10^{-8}]$ & $[\times10^{-8}]$ & [Mm]
\\ \hline\hline\\
A& 1.0000&  1.0000&  0.2241&   -1.68 & -0.25  & -    \\
B& 0.7082&  3.5573&  1.1255&   -0.86 & -11.1  & 30.5 \\
C& 0.1385&  1.1654&  1.8857&   -0.23 & -3.0   & 12.5 \\
D& 1.2807&  6.0726&  1.0625&   -1.15 & -20.6  & 41.8 
\end{tabular}
\caption{Relationship between the horizontal power
distribution at 5Mm depth and the superadiabaticity
of the convection zone. Second and third columns show,
respectively, the total power contained in giant cells
($P_{G}$) and supergranular scales ($P_S$) relative to
that of Case A, with the fourth column showing the ratio
of supergranular to giant cell power. Fifth and sixth
columns display, respectively, the maximal value of the
superadiabaticity parameter
$\epsilon\equiv H_T/\Theta_o d\langle\Theta\rangle/dr$
inside driving regions corresponding to giant cells
($100\text{Mm}<4H_\rho<300\text{Mm}$) and supergranular
scales ($20\text{Mm}< 4H_\rho < 50\text{Mm}$), with
$H_T\equiv -(d\ln T_o/dr)^{-1}$ the temperature scale
height. The last column shows the characteristic depth
of the superadiabatic region below the surface for Cases
B-D (estimated as the smallest depth for which
$\epsilon \le 10^{-8}$).\label{table} \vspace{-0.25cm}
} 
\end{center} 
\end{table}

\section{Summary \& Remarks} \label{sec:rem}
These surface driven convective experiments demonstrate that low
entropy fluid parcels generated in a cooling layer can lead to a
mean thermodynamic state that is strongly superadiabatic in a
narrow region, smoothly transitioning to very nearly adiabatic
stratification below, much more adiabatic than has been achieved
by other simulations to date. The depth of the superadiabatic region
depends on the entropy contrast and density of the downflowing plumes, 
and the convective modes of that layer then determine the 
velocity power spectrum observed~\citep{Lord14}.

Cold fluid parcels generated in the cooling layer transit the
convection zone over a time short compared to numerical diffusion
timescales. Thus the stratification of the upper convection zone
is well approximated by the contribution adiabatically descending
cool parcels make to the mean state strictly by their presence. 
The filling factor of the downflows decreases with the increasing
mean density, yielding, because of the steep stratification, a
nearly adiabatic profile at depth.  The ratio of power at supergranular 
to giant cell scales reflects this, increasing in those simulations
with a shallower transition to adiabatic stratification. 

The short transit time of the cold downflowing plumes across the
solar convection zone compared to the characteristic timescale of
radiative heating in the solar interior suggests that heat transport
is highly non-local (e.g. \cite{Spruit97}). Similar to our simulations,
the change in the mean stratification of the upper solar convection zone 
then reflects the plumes' presence and the decrease in their filling
factor with depth,  as opposed to diffusive processes, which are minimized
in our solutions and likely insignificant in the Sun.  The solar
supergranulation then reflects density scale height  at the depth at
which the solar mean state becomes essentially isentropic.  This is 
quite shallow because of the low density of the granular downflows and
the rapid increase in the mean density of the subphotospheric layers,
increasing by a factor of $\thicksim 1.5\times 10^4$ in the upper 20Mm. 
If this picture is correct, supergranulation represents the largest
buoyantly driven convective scale of the Sun.

\section*{Acknowledgements}
We thank Regner Trampedach, Axel Brandenburg, and Piotr
Smolarkiewicz. This work utilized the Janus supercomputer,
supported by the National Science Foundation (award number
CNS-0821794) and the University of Colorado Boulder. The
Janus supercomputer is a joint effort of the University
of Colorado Boulder, the University of Colorado Denver
and the National Center for Atmospheric Research. J.-F.C.
acknowledges support from the University of Colorado's
George Ellery Hale Postdoctoral Fellowship. M.P.R.'s work
was partially supported by NASA award NNX12AB35G.

\clearpage


\begin{thebibliography}{}
\bibitem[Berrilli et al.(2005)]{Berrilli05} Berrilli, F.,
Del Moro, D., Russo, S., Consolini, G., \& Straus, T.\ 2005,
\apj, 632, 677 
\bibitem[Berrilli et al.(2013)]{Berrilli13} Berrilli, F.,
Scardigli, S., \& Giordano, S.\ 2013, \solphys, 282, 379 
\bibitem[Cattaneo et al.(2001)]{Cat01} Cattaneo, F., Lenz,
D., \& Weiss, N.\ 2001, \apjl, 563, L91 
\bibitem[Christensen-Dalsgaard et al.(1996)]{CD1996}
Christensen-Dalsgaard, J., Dappen, W., Ajukov, S.~V., et al.\ 1996,
Science, 272, 1286 
\bibitem[Cossette et al.(2013)]{Coss13} Cossette, J.-F., 
Charbonneau, P., \& Smolarkiewicz, P.~K.\ 2013, \apjl, 777, L29 
\bibitem[Crouch et al.(2007)]{Crouch07} Crouch, A.~D.,
Charbonneau, P., \& Thibault, K.\ 2007, \apj, 662, 715
\bibitem[Duvall \& Birch(2010)]{Duval10} Duvall, T.~L., Jr.,
\& Birch, A.~C.\ 2010, \apjl, 725, L47-L51 
\bibitem[Featherstone \& Miesch(2015)]{MieschFea15} Featherstone,
N.~A., \& Miesch, M.~S.\ 2015, \apj, 804, 67 
\bibitem[Gastine et al.(2013)]{Gastine13} Gastine, T., Wicht, J., 
\& Aurnou, J.~M.\ 2013, Icarus, 225, 156 
\bibitem[Ghizaru et al.(2010)]{Ghiz10} Ghizaru, M., 
Charbonneau, P., \& Smolarkiewicz, P.~K.\ 2010, \apjl, 715, L133 
\bibitem[Goldbaum et al.(2009)]{Gold09} Goldbaum, N., Rast, M.~P.,
Ermolli, I., Sands, J.~S., \& Berrilli, F.\ 2009, \apj, 707, 67 
\bibitem[Grabowski \& Smolarkiewicz(2002)]{Grab02} Grabowski, W.~W.,
\& Smolarkiewicz, P.~K.\ 2002, Monthly Weather Review, 130, 939
\bibitem[Greer et al.(2015)]{Greer15} Greer, B.~J., Hindman, 
B.~W., Featherstone, N.~A., \& Toomre, J.\ 2015, \apjl, 803, L17 
\bibitem[Hanasoge et al.(2010)]{Hana10} Hanasoge, S.~M., 
Duvall, T.~L., Jr., \& DeRosa, M.~L.\ 2010, \apjl, 712, L98 
\bibitem[Hanasoge et al.(2012)]{Hana12} Hanasoge, S.~M., 
Duvall, T.~L., \& Sreenivasan, K.~R.\ 2012, Proceedings of
\bibitem[Hanasoge et al.(2016)]{Hana16} Hanasoge, S., Gizon, L.,
\& Sreenivasan, K.~R.\ 2016, Annual Review of Fluid Mechanics, 48, 191 
the National Academy of Science, 109, 11928 
\bibitem[Hanasoge \& Sreenivasan(2014)]{Hana14} Hanasoge, S.~M.,
\& Sreenivasan, K.~R.\ 2014, \solphys, 289, 3403 
\bibitem[Hathaway et al.(2000)]{Hatha00} Hathaway, D.~H., Beck, 
J.~G., Bogart, R.~S., et al.\ 2000, \solphys, 193, 299 
\bibitem[Hathaway et al.(2015)]{Hatha15} Hathaway,
D.~H., Teil, T., Norton, A.~A., \& Kitiashvili, I.\ 2015, \apj, 811, 105 
\bibitem[Hathaway et al.(2013)]{Hatha13} Hathaway, D.~H., 
Upton, L., \& Colegrove, O.\ 2013, Science, 342, 1217 
\bibitem[Hotta et al.(2015)]{Hotta14} Hotta, H., Rempel, M., 
\& Yokoyama, T.\ 2015, \apj, 803, 42 
\bibitem[Leighton et al.(1962)]{Leigh62} Leighton, R.~B., 
Noyes, R.~W., \& Simon, G.~W.\ 1962, \apj, 135, 474 
\bibitem[Leitzinger et al.(2005)]{Leit05} Leitzinger, M.,
Brandt, P.~N., Hanslmeier, A., P{\"o}tzi, W., \& Hirzberger,
J.\ 2005, \aap, 444, 245
\bibitem[Lipps  \& Hemler(1982)]{LH82}
Lipps, F.~B., \& Hemler, R.~S.\ 1982, Journal of Atmospheric
Sciences, 39, 2192
\bibitem[Lord et al.(2014)]{Lord14} Lord, J.~W., Cameron, 
R.~H., Rast, M.~P., Rempel, M., \& Roudier, T.\ 2014, \apj, 793, 24 
\bibitem[Miesch(2005)]{Miesch05} Miesch, M.~S.\ 2005, Living 
Reviews in Solar Physics, 2, 1 
\bibitem[Miesch et al.(2008)]{Miesch08} Miesch, M.~S., Brun, 
A.~S., De Rosa, M.~L., \& Toomre, J.\ 2008, \apj, 673, 557 
\bibitem[Miesch \& Toomre(2009)]{MieschToomre09} Miesch, M.~S.,
\& Toomre, J.\ 2009, Annual Review of Fluid Mechanics, 41, 317 
\bibitem[Nordlund et al.(2009)]{Nord09} Nordlund, {\AA}., 
Stein, R.~F., \& Asplund, M.\ 2009, Living Reviews in Solar Physics, 6, 2 
\bibitem[November et al.(1981)]{Novem81} November, L.~J., 
Toomre, J., Gebbie, K.~B., \& Simon, G.~W.\ 1981, \apjl, 245, L123 
\bibitem[November(1989)]{Nov89} November, L.~J.\ 1989, \apj, 344, 494 
\bibitem[Prusa \& Smolarkiewicz(2003)]{Prusa03} Prusa, M.~P.,
\& Smolarkiewicz, P.~K. \ 2003, Journal of Computational Physics, 190, 601
\bibitem[Prusa et al.(2008)]{Prusa08} Prusa, J.~M., Smolarkiewicz, P.~K.,
\& Wyszogrodzki, A.~A.\ 2008, Comput. Fluids, 37, 1193
\bibitem[Racine et al.(2011)]{Rac11} Racine, {\'E}., 
Charbonneau, P., Ghizaru, M., Bouchat, A., 
\& Smolarkiewicz, P.~K.\ 2011, \apj, 735, 46 
\bibitem[Rast(2003)]{Rast03} Rast, M.~P.\ 2003, \apj, 597, 1200 
\bibitem[Rast \& Toomre(1993)]{RastToomre93} Rast, M.~P., \&
Toomre, J.\ 1993, \apj, 419, 224 
\bibitem[Rincon(2007)]{Rincon07} Rincon, F.\ 2007, IAU 
Symposium, 239, 58 
\bibitem[Rieutord et al.(2000)]{Rieutord00} Rieutord, M.,
Roudier, T., Malherbe, J.~M., \& Rincon, F.\ 2000, \aap, 357, 1063 
\bibitem[Roudier et al.(2012)]{Roudier12} Roudier, T., Rieutord, M.,
Malherbe, J.~M., et al.\ 2012, \aap, 540, A88
\bibitem[Simon \& Leighton(1964)]{SimLeigh64} Simon, G.~W.,
\& Leighton, R.~B.\ 1964, \apj, 140, 1120 
\bibitem[Smolarkiewicz et al.(2001)]{Smo01} Smolarkiewicz,  P.~K.,
Margolin, L.~G., \& Wyszogrodzki, A.~A.\ 2001, Journal of Atmospheric
Sciences, 58, 349
\bibitem[Smolarkiewicz(2006)]{Smolar06} Smolarkiewicz, P.~K.\
2006, International Journal for Numerical Methods in Fluids, 50, 1123
\bibitem[Smolarkiewicz \& Charbonneau(2013)]{Smolar13} Smolarkiewicz, P.~K.,
\& Charbonneau, P.\ 2013, Journal of Computational Physics, 236, 608
\bibitem[Smolarkiewicz \& Prusa(2002)]{SmolarPrusa02} Smolarkiewicz, P.~K.,
\& Prusa, J.~M.\ 2002, International Journal for Numerical Methods in Fluids,
39, 799
\bibitem[Spruit(1997)]{Spruit97} Spruit, H.\ 1997, \memsai, 68, 397 
\bibitem[Thompson et al.(2003)]{Thompson03} Thompson, M.~J.,
Christensen-Dalsgaard, J., Miesch, M.~S., \& Toomre, J.\ 2003,
\araa, 41, 599 
\bibitem[Warn-Varnas et al.(2007)]{Warn07} Warn-Varnas, A.,
Hawkins, J., Smolarkiewicz, P.~K., et al.\ 2007, Ocean Modelling, 18, 97

\end{thebibliography}
\end{document}